\documentclass[cup5b]{caps}
\usepackage{graphicx}
\usepackage{amssymb}
\usepackage{ociwsymp3e}  

\HeadText{B. R. McNamara}  

\def\mathfont#1{\ifmmode{#1}\else{$#1$}\fi} 
\def\lae{\mathrel{<\kern-1.0em\lower0.9ex\hbox{$\sim$}}}  
\def\gae{\mathrel{>\kern-1.0em\lower0.9ex\hbox{$\sim$}}}  

\begin{document}

\pagenumbering{arabic}

\author[]{B. R. McNamara\\Department of Physics \& Astronomy, Ohio University,
Athens, OH}

%
%

\chapter{Magnetic Bubbles in Galaxy Clusters}

\begin{abstract}

I discuss {\it Chandra} X-ray Observatory measurements
of cavities in galaxy clusters and their implications for 
heating the intracluster gas.  The emerging paradigm for 
cooling flows has important implications for understanding
self-regulated galaxy formation.  

\end{abstract}

\section{Introduction}

{\it Chandra} X-ray images of galaxy cluster cores have revealed
a wealth of structure. Once thought to be relatively smooth, quiescent
environments, cluster cores are now known to be 
dynamically complex regions of the universe.  Examples of commonly
observed structures include sharp surface brightness 
edges associated with mergers (Markevitch et al. 2000),
filaments associated with cooling or dynamical wakes (Fabian et al. 2001),
and the topic of this discussion: 
cavities or bubbles created by interactions between
radio sources and the hot gas surrounding them (McNamara et al. 2000).

\section{Properties of Cavities in Clusters}

Cavities have been identified in at least a dozen clusters 
over the past three
years (B\^{\i}rzan et al. 2004).   The archetypes, Hydra A (McNamara
et al. 2000) and Perseus (B\"ohringer et al. 1993, Fabian et al. 2000),
are typical of most systems: 
twin surface brightness depressions $10-20$ 
kpc in diameter lying at distances of $10-30$ kpc from the nucleus of the cD.
The cavities are devoid of thermal gas at the characteristic temperature 
of their surroundings.  They are filled, however, with radio-synchrotron
emitting particles and their accompanying magnetic
field.  The equipartition pressures within the cavities are generally 
between $5-10$ times lower than their surroundings 
(Nulsen et al. 2002, Blanton et al. 2001), which implies that they
are short lived.  Their ubiquity and and apparently advanced
ages suggest otherwise.  

Cavity pairs harboring faint radio 
emission located well beyond
their brighter central radio sources, dubbed ``ghost cavities'' 
have been discovered in several clusters 
(eg., Fabian et al. 2000, McNamara et al. 2001).  
Propelled by buoyant forces, the time required for these cavities to rise 
to their current locations can
approach $\sim 10^8$ yr, or several dynamical timescales. 
The cavities can survive against collapse 
to such an advanced age only if they are supported 
by internal pressure and if they are hydrodynamically stable.   
The source of this internal pressure is
unknown. Candidates include a hot, dilute thermal
plasma, a relativistic gas, or magnetic field.  The temperature
a thermal gas must have in order to support the cavities 
against collapse, and at the
same time, elude detection in existing {\it Chandra} images is
$>15$ keV (Blanton et al. 2001, Nulsen et al. 2002).  Synchrotron
emission limits seem to preclude magnetic field as the sole 
source of internal pressure in the Perseus cavities (Fabian et al. 2002),
so they probably contain very hot, possibly relativistic, gas.
Their relatively stable configurations must be aided
by surface tension in the bubble rims.  Otherwise, 
the bubbles would  quickly disintegrate by
Rayleigh-Taylor forces (Soker, Blanton \& Sarazin 2002).
Magnetic fields are a likely source of surface tension (De Young 2003).

That the temperatures of the rims are as cool or 
cooler than the ambient gas was perhaps {Chandra}'s most
surprising discovery (McNamara et al. 2000, Nulsen et al. 2002).  
Early theoretical models incorrectly predicted hot rims associated with 
shock fronts (Heinz, Reynolds, \& Begelman 1998).  
The cool rims imply that the cavities behave
like bubbles (Churazov et al. 2001), rising buoyantly at or below the
sound speed.  The cool gas along the rims was probably lifted by the
bubbles from the cooler central regions of the clusters 
(Nulsen et al. 2002, Blanton
et al. 2001). This general picture has been bolstered by the recent discovery 
of sound ``ripples'' emanating from the two inner
cavities in the Perseus cluster (Fabian et al. 2003). 

\section{Cavity Demographics \& Energetics}

Cavities have been observed in giant elliptical galaxies,
such as M84 (Finoguenov \& Jones 2001), groups, such as HGG 62,
and clusters (B\^{\i}rzan et al. 2004).  Their energy content
ranges between $pV\sim 10^{55}~{\rm erg ~ s^{-1}}$ in
isolated galaxies and groups to $\sim 10^{59}~{\rm erg ~ s^{-1}}$
in rich clusters; their ages range between 
$\sim 10^7~{\rm yr}-10^8~{\rm yr}$.  The total energy associated with
the cavities can be four times this number if they are filled with 
a relativistic gas.

\section{Can Magnetic Bubbles Quench Cooling flows?}

The persistent symptom of the so-called cooling flow problem has been that  
the cooling rates exceed the
star formation rates by at least an order of magnitude.  
This situation has changed dramatically in recent years.  
New XMM-Newton and {\it Chandra} observations (Peterson et al. 2001) 
have placed limits on cooling to low temperatures that are factors of
$5-10$ below the old {\it Einstein} and {\it Rosat} rates.
This reduction implies that the gas is being maintained at keV
temperatures by a persistent energy source.  Magnetic bubbles
are a plausible source of this energy.

There is growing evidence that bubbles are produced periodically
in cooling flow clusters.  The older and radially distant ghost 
cavities in the
Perseus cluster (Fabian et al. 2000) and the Abell 2597 cluster 
(McNamara et al. 2001) were probably created by an earlier generation of the central radio
source.  The ghost cavities are associated with radio lobes 
that have since detached from their jets and have traveled
buoyantly to their current locations over the past $\sim 10^8$ yr.
In the mean time, the rejuvenated central radio source
has created a new set
of radio-filled cavities near the nucleus.  These and perhaps
other systems launch cavities 
every several tens of Myr.  Coupled with the fact that cD galaxies
in cooling flows are radio audible $\sim 70\%$ of the time (Burns 1990),
the rising bubbles may deposit up to $\sim 10^{61}~{\rm erg}$ of energy
into the intracluster medium over their lives (McNamara et al. 2001).  
This would be enough energy to impede or
quench a moderately sized cooling flow.  The production rate required to
prevent cooling in the
Perseus cluster, for example, is one bubble pair every $\sim 10^7$ yr (Fabian
et al. 2003).  
 
Bubble production may be able to retard or quench cooling in many systems,
but apparently not throughout the lives of all systems.
For example, the Abell 1068 cluster harbors moderate 
cooling at a rate $\lae 140~{\rm M_{\odot}~yr^{-1}}$ (Wise et al. 2004).  
The star formation rate in its cD galaxy 
is $\sim 70 ~{\rm M_{\odot}~yr^{-1}}$.  To within their uncertainties,
the cooling and star formation rates are consistent with each other
(McNamara et al. 2004), and there is no need to appeal to heating.
Furthermore, Abell 1068 has no cavities,
its radio source is weak, conduction is too inefficient to prevent
cooling, and supernovae associated with the starburst are incapable
of quenching the cooling flow.  Abell 1068 has the qualities of
a classical cooling flow, at least at this stage of its life.
Therefore, all cooling flows do not achieve a steady
balance between heating and cooling throughout their lives.

\section{Conclusions \& Speculations about a New Cooling Flow Paradigm}

Cooling flows are usually messy systems.  Even those in
which bubble production is energetically sufficient to prevent cooling,
cold gas and young stars abound.  
Cooling to low temperatures is probably occurring
within cD galaxies along filaments of cool gas located near 
the sites of star formation (McNamara et. al. 2000, Blanton et al. 2003,
McNamara et al. 2004).  The star formation itself often occurs
in bursts.  

The emerging cooling flow paradigm no longer supports the notion of long-term,
steady cooling.  Instead, a cooling cycle that fuels
repeated episodes of star formation is established, followed 
by accretion onto the central black hole.  A radio outburst
ensues, creating bubbles that reheat the cooling gas.
This cycle repeats. Thermal conduction may 
play a critical role in maintaining the feedback loop 
(Ruszkowski \& Begelman 2002).
The existence of a trend between the central X-ray luminosity and cavity energy
(B\^{\i}rzan et al. 2004) suggests that this process proceeds in a 
self-regulatory fashion.

This primitive sketch of a cooling flow must include the essential physics 
of self-regulated  galaxy formation.  In this picture, 
black holes regulate the rate at which bulges form. 
Similar processes may have been operating during the earliest phases
of galaxy formation when the relationship between
black hole mass and bulge velocity dispersion was established
(Fererese \& Merritt 2000, Gebhardt et al. 2000). 
In addition, bubble production is a potential source of 
preheating during the construction phases 
of groups and clusters.

\section{Acknowledgements}

I thank my collaborators Michael Wise, Paul Nulsen, Liz Blanton, and Craig
Sarazin, and my students Laura B\^{\i}rzan and David Rafferty.  
This research
was supported by NASA Long Term Space Astrophysics Grant NAG5-11025, 
Chandra Archival Research Grant AR2-3007X, and a grant from the
Department of Energy through the Los Alamos National Laboratory.

\end{document}